\documentstyle[12pt]{article}
\newcommand{\doublespace}{\renewcommand{\baselinestretch}{1.75}
   \Large\normalsize}

\renewcommand{\ref}[1]{\raisebox{.6ex}{[#1]}}

\newcommand{\be}{\begin{equation}}
\newcommand{\ee}{\end{equation}}
\newcommand{\ba}{\begin{array}}
\newcommand{\ea}{\end{array}}

\begin{document}

\doublespace

\title{ Spectral Flow, the Magnus Force, and the Josephson-Anderson Relation  }

\author{P. Ao$^{\ast}$   \\ Department of Theoretical Physics                 \\
Ume\aa{\ }University, S-901 87, Ume\aa, SWEDEN  \\ }

\maketitle

\begin{abstract}
We show that the spectral flow due to a moving vortex is 
identical to the phase slippage process,
and 
conclude that its evaluation confirms the results by the
Berry's phase calculation of the Magnus force.
\end{abstract}

\newpage

Consider a rectilinear vortex moving in a superconductor  
with a small velocity ${\bf v}_v$
relative to the background of the crystal lattice,
both the Berry's phase\cite{ao} and the direct total transverse force\cite{thouless} 
calculations lead to the Magnus force:
\be
   {\bf F}_M =-L\; q_v h \frac{\rho_s}{2} \; {\bf v}_v \times \hat{\bf z} \; .
\ee
Here $\rho_s$ is the superfluid electron number density, $q_v = \pm 1$ is the
vorticity, and the vortex line is along the $z$-direction with the length $L$.
The crystal lattice background of the superconductor 
is held at rest in the laboratory frame to avoid 
further complications.
Looking from the point of view of the electron fluid, a Fermi liquid,
moving vortices cause phase slippages, generating a potential
difference. Therefore the electron fluid also feels a force, corresponding to the
measured electric field. This process has been analyzed
by Josephson\cite{josephson} and Anderson\cite{anderson}.
Recently, it has been reanalyzed by the calculation of  the
spectral or momentum flow in the electron fluid\cite{volovik,gaitan}.
Since this spectral flow has a sign opposite to that of the Magnus force, 
one might be tempted to conclude 
that it cancels the Magnus force.
The purpose of the present letter is to show 
that those two views of looking at the consequences of a moving vortex 
are equivalent as the action-reaction forces. 
This equivalence must have been implicitly
contained in the literature, but it has not been explicitly stated yet.
Instead, some confusion still exists.\cite{volovik2,makhlin}

We start by studying the momentum flow in the electron fluid due to a moving
vortex trapped by a potential located at 
${\bf r}_0$. The trapping potential can be arbitrarily weak, and just
defines the vortex position. 
The electron fluid system is homogeneous other than this trapping potential,
and the system of vortex-electron fluid is translational invariant. 
The momentum flow in the electron fluid due to 
the moving of the trapping potential, the moving of the vortex, is
\be
  \frac{d }{d t} {\bf P}(t) 
           = Tr [ \dot{\hat{\rho} }(t) \hat{\bf P} ] \; .
\ee
Here the  density matrix for the whole electron fluid is governed by the equation
$
   i\hbar \dot{\hat{\rho} }(t) = [ \hat{H}, \hat{\rho}(t)]
$,
with $\hat{\bf P}$ the total momentum operator and $\hat{H}$ 
the truncated Hamiltonian of the system: the lattice phonon induced 
effective electron-electron interaction 
is treated as an attractive one, which gives arise to the superconductivity. 
One may include the lattice dynamics into the formulation. 
In this case, because the electron-lattice
system consists of a clean superconductor, our following conclusion will remain unmodified.
We first evaluate the momentum flow from a `safe' distance far away from the
vortex core. 
Later we point out that it can be evaluated near the vortex
core by the spectral flow method and the results of the two calculations 
are equal as guaranteed by the momentum conservation law.
We note that for a slow moving vortex, the adiabatic condition holds: 
At a given time the density matrix $\hat{\rho}$ of the electron fluid system 
can be approximated by its instantaneous equilibrium density matrix 
$\hat{\rho}_0$ and the deviation $\hat{\rho}_1$: 
$ \hat{\rho} = \hat{\rho}_0 + \hat{\rho}_1$.
Here $\hat{\rho}_1$ is determined by the equation
$
  [ \hat{H}, \hat{\rho}_1] = i\hbar \dot{\hat{\rho} }_0 
$,
and   
\be
   \hat{\rho}_0(t) = \sum_n f_n |\Psi_n >< \Psi_n | \; ,
\ee
with $f_n = e^{ - E_n/k_B T}/\sum_n e^{ - E_n/k_B T}$ the normalized Boltzmann
factor and $\Psi_n$ the n-th eigenstate of the Hamiltonian.
The eigenenergies $\{ E_n \}$ are independent of time. 
The time dependence of the
density matrix is through the vortex position dependence of the wavefunction.
Then we have, to the lowest order in the vortex velocity,
\be
   \frac{d }{d t} {\bf P}(t) = 
     Tr \left\{  \sum_n  f_n {\bf v}_v \cdot 
     [ |\nabla_{{\bf r}_0} \Psi_n >< \Psi_n | + 
      |\Psi_n >< \nabla_{{\bf r}_0} \Psi_n | ] \hat{\bf P} \right\} \; .
\ee
Here the operator $\nabla_{{\bf r}_0}$ is the gradient with respect to the
vortex position, and ${\bf v}_v = \dot{\bf r}_0$.
Since the vortex position, or the position of the trapping potential,
is the only reference point in the electron fluid system, we have the identity
\be
   \nabla_{{\bf r}_0} \Psi_n(\{ {\bf r}_j \},{\bf r}_0)  
   = - \sum_{j} \nabla_{{\bf r}_j} \Psi_n(\{ {\bf r}_j \}, {\bf r}_0) \; ,
\ee
with ${\bf r}_j$ the position of the j-th electron in the system.
Note that the total momentum operator is 
$\hat{\bf P}
 = - i \hbar \sum_{j} \nabla_{{\bf r}_j}\equiv  - i \hbar \nabla_{\bf R} $,
Eq.(4) gives us
\[
  \frac{d }{d t} {\bf P}(t) = i\hbar  
   \left\{  \sum_n  f_n 
     [ - < \nabla_{\bf R} \Psi_n |{\bf v}_v \cdot\nabla_{\bf R} \Psi_n > + 
     < {\bf v}_v \cdot \nabla_{\bf R} \Psi_n | \nabla_{\bf R} \Psi_n > ]
       \right\} 
\]
\be
   = - i\hbar {\bf v}_v \times
   \sum_n  f_n \int \prod_j d^3{\bf r}_j 
     [ \nabla_{\bf R} \times \nabla_{{\bf R}'} 
        \Psi_n(\{{\bf r}'_j\})  \Psi_n^{*}(\{ {\bf r}_j\}) ]
           |_{ \{ {\bf r}'_j \}= \{ {\bf r}_j \} } \; .
\ee
Following the same procedure as in Ref.\cite{thouless}, we first reduce the
N-body density to the one-body density matrix $\rho_1$:
\be
    \frac{d }{d t} {\bf P}(t) = - i\hbar {\bf v}_v \times
    \int d^3{\bf r}_1 
     [ \nabla_{{\bf r}_1} \times \nabla_{{\bf r}'_1 } 
   \rho_1( {\bf r}'_1; {\bf r}_1 ) ]|_{{\bf r}'_1 = {\bf r}_1 } \; ,
\ee
with 
\be
  \rho_1( {\bf r}'_1; {\bf r}_1 ) \equiv N
   \sum_n  f_n  \int \prod_{j\neq 1} d^3{\bf r}_j 
    \Psi_n({\bf r}'_1,\{{\bf r}_j\})\Psi_n^{*}({\bf r}_1, \{{\bf r}_j\})  \; ,
\ee
and $N$ the total electron number.
Then we use the Stokes' theorem
to evaluate the integral far away from the vortex core, and obtain
\be
   \frac{d }{d t} {\bf P}(t) = L \; \hbar 
      {\bf v}_v \times \; \hat{\bf z} \oint d{\bf r}
      \cdot \left.\frac{i}{2}[(\nabla_{\bf r} - \nabla_{{\bf r}'} )
        \rho_1({\bf r}'; {\bf r}) ] \right|_{ {\bf r}'={\bf r} }
     = L \; q_v h \frac{\rho_s}{2} 
          {\bf v}_v \times \hat{\bf z} \; ,
\ee
which shows that the force felt by the electron fluid has 
the same magnitude as the Magnus force,  but with the opposite sign.
We will return to this point below.
In the integration leading to Eq.(9) we have used the fact that the integrand
is the momentum density. 
The two-fluid model has been employed to account for
the fact that the momentum generated by the vortex corresponds to
 a supercurrent. 
The momentum flow calculation can also be performed near the vortex core 
by counting the flow of the energy spectrum, 
the so-called spectral flow method,
and have been done in Ref.\cite{volovik}.
The main idea is that, 
by linearizing the time-dependent Bogoliubov-de Gennes equation near the Fermi
surface, a pseudorelativistic Dirac equation will be obtained.
This linearization reduces the original 3+1 dimensional problem to 
an effective 1+1 dimensional one.
As a vortex moves, there is a continuous flow of the energy spectrum,
emerging from (or sinking into) the Fermi sea.
The spectral flow rate is proportional to the vortex velocity.
By counting their contributions to the momentum, 
the same result as Eq.(9) has been arrived at in Ref.\cite{volovik}. 
This procedure has been further 
verified by a model calculation.\cite{makhlin}
The agreement between two seemingly totally 
different ways, looking
from a safe distance away from the core and watching the spectral
flow near the core,  of calculating the momentum flow 
may first appear surprising. 
It has been, however, explicitly demonstrated in Ref.\cite{stone} 
in a slightly different context in the $^3$He-A phase 
that they are indeed the complementary 
two ways of keeping track of the same physics, and they are equal
as a result of the momentum conservation in the electron fluid system.
In the present situation the link between those two ways
has been discussed in Ref.\cite{gaitan}.  
It is the differential form of the momentum flow:
\be
   \frac{\partial }{\partial t} ( m^{*}{\bf v}_s ) + \nabla \mu = 
     q_v h {\bf v}_v \times \hat{\bf z} \; \delta^2({\bf r} - {\bf r}_0) \; ,
\ee
with the electrochemical potential $\mu$ as the sum of 
the chemical potential $\mu_0$, 
the fluid kinetic energy and the electric potential
 $\mu = \mu_0 + \frac{m^{*}}{2} v_s^2 + e^{*} ( \phi + \frac{1}{c} 
 \frac{\partial }{\partial t  } 
      \int^{\bf r} d{\bf r}'\cdot {\bf A}({\bf r}',t)) $,
and the superfluid velocity distribution $m^{*}{\bf v}_s 
= \hbar\nabla\theta - \frac{e^{*}}{c} {\bf A} $.
Here $m^{*}$ and  $e^{*}$ are effective mass and charge of a Cooper pair,
respectively, and $\theta$ is the phase of the superconducting 
condensate wavefunction.
Eq.(10) is gauge invariant, and is a precise 
statement about the phase slippage process due to a moving vortex. 
Its various consequences have been explored by 
Josephson\cite{josephson} and Anderson\cite{anderson}. An identical
equation for the special case of 
${\bf v}_v = {\bf v}_s$ in the neutral superfluid has been studied
by Anderson\cite{anderson}. We may call Eq.(10) the
Josephson-Anderson relation.

It is clear now that a moving vortex feels a transverse force,
the Magnus force; if one looks from the point of view of the electron fluid, 
the fluid also feels a force with the  
magnitude equal to but the sign opposite to the Magnus force. 
The two forces are acting on two different objects, 
the vortex and the electron fluid, and are the action-reaction forces.
In the following we strengthen this point by a straightforward proof.

We write Eq.(2) in its equivalent form of the Ehrenfest theorem:
\be
     \frac{d }{d t} {\bf P}(t) 
      = - Tr [\hat{\rho}(t) \nabla_{\bf R} \hat{H} ] \; .
\ee
Using $\nabla_{\bf R} \hat{H} = - \nabla_{{\bf r}_0} \hat{H} $,
we obtain
\be
     \frac{d }{d t} {\bf P}(t) 
       = Tr [\hat{\rho}(t) \nabla_{{\bf r}_0} \hat{H} ] = - {\bf F}_M \; .
\ee
This shows that 
the two forces under discussion are indeed the action-reaction forces.
In the last equality we have used the formal definition of the 
Magnus force\cite{thouless}.

The present result, the equivalence between the Magnus force 
and the spectral flow as action-reaction forces, may seem obvious.
Nevertheless it has never been explicitly spelled out in the literature.
Instead, some recent work have treated the spectral flow 
as a way to cancel the Magnus force\cite{volovik,volovik2,makhlin},
which is incorrect according to the present demonstration.
It should be pointed out that the spectral flow is a counting of
contributions from extended states. 
There is no involvement of the localized core states.
This can be explicitly checked by expressing the one-body density matrix in 
Eq.(9) in terms of extended and localized states, and latter
gives zero contribution, as having been noted in Ref.\cite{thouless}
in the evaluation of the Magnus force. 
Incidentally, in Ref.\cite{volovik} the Wess-Zumino term for a moving
vortex has been identified as the same force as the one 
due to the spectral flow. 
Recent as well as earlier work have shown that the Wess-Zumino term 
gives exactly 
the Magnus force\cite{hatsuda,aatz,stone2,gaitan}, not the spectral flow.  

Comparing  the Berry's phase calculation away from the vortex core with
the spectral or momentum flow counting near the vortex core, 
we find that the former only depends on a few global properties of a
superconductor, namely the topology of a vortex, 
and the latter is a rather detailed calculation.
The topological constraints behave like conservations laws.
Results obtained under them should, and have to, be borne
out by detailed calculations, which are concrete realizations.
For the Magnus force, it is indeed the case.

\noindent
{\bf Acknowledgements: } 
The author thanks David Thouless and Qian Niu for numerous discussions, 
and  Mike Stone and Frank Gaitan for 
informative correspondences.
The paper was initiated 
at the Institute of Scientific Information at 
Turin in the fall of 1993, and was shaped into the present form at
the Aspen Center for Physics in the summer of 1995. Their hospitalities
are gratefully acknowledged.
The work was supported in part by Swedish Natural Science Research Council
and by US NSF Grant No. DMR-9220733.

{\ }

\noindent
$^{\ast}$Present address: \\
Department of Physics, Box 351560, University of Washington, Seattle, 
WA 98195, USA.


\begin{thebibliography}{99}

\bibitem{ao}
 P. Ao and D.J. Thouless, Phys. Rev. Lett. {\bf 70}, 2158 (1993); \\
 P. Ao, Q. Niu, and D.J. Thouless, Physica {\bf B194-196}, 1453 (1994). 
\bibitem{thouless} 
 D.J. Thouless, P. Ao, and Q. Niu, {\it Transverse force on a quantized vortex
  in a superfluid}, preprint, High-Tc Update, Nov. 15, 1995.  
\bibitem{josephson}
 B.D. Josephson, Phys. Lett. {\bf 1}, 251 (1962);
                {\it ibid}. {\bf 16}, 242 (1965). 
\bibitem{anderson}
 P.W. Anderson, Rev. Mod. Phys. {\bf 38}, 298 (1966).  
\bibitem{volovik}
 G.E. Volovik, JETP Lett. {\bf 57}, 244 (1993); 
                     JETP {\bf 77}, 435 (1993). 
\bibitem{gaitan}
 F. Gaitan,  J. Phys. Cond. Matt. {\bf 7}, L165 (1995); 
           Phys. Rev. {\bf B51}, 9061 (1995).     
\bibitem{volovik2}
 G.E. Volovik, JETP Lett. {\bf 62}, 65 (1995). 
\bibitem{makhlin}
 Y.G. Makhlin and T.S. Misirpashaev, JETP Lett. {\bf 62}, 83 (1995).  
\bibitem{stone}
 M. Stone and F. Gaitan, Ann. Phys.(N.Y.) {\bf 178}, 89 (1987). 
\bibitem{hatsuda}
 M. Hatsuda, S. Yahikozawa, P. Ao, and D.J. Thouless, 
          Phys. Rev. {\bf B49}, 15870 (1994); and references therein. 
\bibitem{aatz}
 P. Ao, D.J. Thouless, and X.-M. Zhu, Mod. Phys. Lett. {\bf B9}, 755 (1995);\\
 I.J.R. Aitchison, P. Ao, D.J. Thouless, and X.-M. Zhu, Phys, Rev, {\bf B51},
               6531 (1995). 
\bibitem{stone2}
 M. Stone, Int. J. Mod. Phys. {\bf B9}, 1359 (1995). 
\end{thebibliography}
\end{document}